\documentstyle[preprint,osa,eqsecnum]{revtex}

\newcommand\beq{\begin{equation}}
\newcommand\eeq{\end{equation}}
\def\beqa{\begin{eqnarray}}
\def\eeqa{\end{eqnarray}}
\def\bega{\begin{array}}
\def\enda{\end{array}}

\def\vb{\vphantom{|}}

\def\a{\alpha}
\def\b{\beta}

\def\th{{\theta}}

\def\d{\partial}

\def\From{{From}}
\begin{document}
\preprint{Submitted to C.R. Acad. Sc. Paris}
\title{ Geometrical Well Posed Systems for the Einstein Equations}

\author{Yvonne Choquet-Bruhat
\thanks{Electronic address: choquet@circp.jussieu.fr}\\
Gravitation et Cosmologie Relativiste\\
Universit\'e Paris VI, t. 22-12\\
Paris 75252 France\\
and\\
James W. York\thanks{Electronic address: york@physics.unc.edu}\\
Department of Physics and Astronomy\\
University of North Carolina\\
Chapel Hill NC 27599-3255 USA}
\date{June 26, 1995}

\maketitle

\vspace{-12.8cm}
\hfill IFP-UNC-509, TAR-UNC-047
\vspace{11.5cm}

\preprint{Submitted to C.R. Acad. Sc. Paris}

\begin{abstract}
We show that, given an arbitrary shift, the lapse $N$ can be chosen so
that the extrinsic curvature $K$ of the space slices with metric
$\overline g$ in arbitrary coordinates of a solution of Einstein's
equations satisfies a quasi-linear wave equation.  We give a geometric
first order symmetric hyperbolic system verified in vacuum by $\overline
g$, $K$ and $N$.  We show that one can also obtain a quasi-linear wave
equation for $K$ by requiring $N$ to satisfy at each time an elliptic
equation which fixes the value of the mean extrinsic curvature of the
space slices.
\end{abstract}
\newpage

\subsubsection*{SYSTEMES GEOMETRIQUES BIEN POSES POUR LES EQUATIONS
D'EINSTEIN.}

{\bf R\'esum\'e} Nous donnons deux conditions diff\'erentes sur le lapse
pour que la courbure extrins\`eque $K$ des sections d'espace d'une
solution des \'equations d'Einstein satisfasse une \'equation d'onde
quasi-lin\'eaire, le shift \'etant arbitraire.
\vspace{0.5cm}

\subsection*{Version fran\c caise abr\'eg\'ee}

Nous consid\'erons les \'equations d'Einstein sur un espace temps de
dimension 4, de topologie $M\times R$ dont nous \'ecrivons la m\'etrique
$$ ds^2 = -N^2 (\th^0)^2 + g_{ij} \th^i \th^j $$
dans le corep\`ere donn\'e par ($x^i$ coordonn\'ees locales sur $M$,
$t\in R$)
$$ \th^0 = dt,\quad\quad \th^i = dx^i + \b^i dt . $$
Nous d\'efinissons l'op\'erateur $\hat\d_0$ sur les tenseurs d'espace
d\'ependant du temps par (${\cal L}$ est la d\'eriv\'ee de Lie,
$\partial_0$ la d\'eriv\'ee pfaffienne)
$$ \hat\d_0 = \d/\d t -{\cal L}_\b,\quad{\rm alors~~que}\quad \d_0=\d/\d t
- \b^i \d/\d x^i.$$
La courbure extrins\'eque $K$ des sections d'espace $M\times\{t\}$ est
telle que
$$ \hat\d_0 g_{ij} = -2N K_{ij} . $$
Nous utilisons la d\'ecomposition 3+1 du tenseur de Ricci de l'espace
temps pour obtenir l'identit\'e,
(2.1) donn\'ee au \S2, g\'en\'eralisation de celle
donn\'ee dans [2] dans le cas $\beta=0$, o\`u on a pos\'e
$H=K^i_i$ et $(ij)=ij+ji$.
Cette identit\'e donne une \'equation d'ondes pour $K$ dans les cas
suivants:
\begin{enumerate}
\item $N$ est choisi tel que $\d_0 N + N^2 H=0$.

\item $N$ est choisi tel que $H=h$, une fonction donn\'ee.
\end{enumerate}
Nous donnons dans le cas 1 un syst\`eme sym\'etrique hyperbolique du
premier ordre satisfait par les espaces temps einsteiniens.

\section*{Introduction}

We examine the Cauchy problem \cite{CBY} for general relativity as the
time history of the geometry of a spacelike hypersurface.  Constraints on
the initial data, $\overline g$, the space metric, and $K$, the
extrinsic curvature, can be posed and solved as an elliptic system by
known methods.  However, the equations of motion for $\overline g$ and
$K$, which are essentially the Arnowitt-Deser-Misner canonical equations,
do not, despite their usefulness, manifest mathematically the physical
propagation of gravitational effects along the light cone.  These
equations, of course, contain gauge effects and cannot, therefore, yield
directly a physical wave equation, that is, a hyperbolic system with
suitable characteristics.  (For a review of the relevant geometry, see,
for example, \cite{Yor}.)

In this paper we give two different methods for obtaining exact nonlinear
physical wave equations from the equations of motion of $\overline g$ and
$K$.  One of these, which relies on a ``harmonic'' time-slicing
condition, gives equations of motion completely equivalent to a
first-order symmetric hyperbolic system with only physically appropriate
characteristics.  We construct this system explicitly.  Among the
propagated quantities is, in effect, the Riemann curvature.  The space
coordinates and shift vector are arbitrary; and, in this sense, the
system is gauge-invariant.

Our gauge invariant nonlinear hyperbolic system, being {\it exact and
always on the physical light cone}, is well suited for use in a number of
problems that now confront gravity theorists.  These include large-scale
computations of astrophysically significant processes (such as black hole
collisions) that require efficient stable numerical integration,
extraction of gravitational radiation with arbitrarily high accuracy from
Cauchy data, gauge-invariant perturbation and approximation methods, and
posing boundary conditions compatibly with the causal structure of
spacetime.

\section{NOTATIONS AND 3+1 DECOMPOSITION}

We consider a manifold $V$ of dimension 4, which has the topology
$M\times R$, with $M$ a ${\cal C}^\infty$, 3 dimensional manifold.  We
denote by $(x,t)\in M\times R$ a point of $V$.  When we take local
coordinates they will always be adapted to the product structure: $x^{i}$,
$i=1,2,3$ are coordinates on $M$ and $x^0\equiv t\in R$.  We consider on $V$
a pseudo-riemannian metric $g$ of lorentzian signature $(-+++)$.  It
induces a properly riemannian metric $\bar g_t$ on each space submanifold
$M_t\equiv M\times \{t\}$.  We denote by $\beta$ and $N$ respectively the
shift and the lapse of the foliation.  We define a coframe $\th^{\alpha}$
by
$$ \th^0 \equiv dt,\quad\quad \th^{i} \equiv dx^{i} +\b^{i} d t. $$
The corresponding Pfaff derivatives $\d_\alpha$ are
$$ \d_0 \equiv {\d\over \d t} - \b^i \d_i, \quad\quad \d_i \equiv
{\d\over \d x^{i}}. $$
In this coframe the metric reads
$$ ds^2 \equiv g_{\a\b} \th^{\a}\th^{\b} \equiv -N^2 (\th^0)^2 + g_{ij}
\th^i \th^j. $$
We define the operator $\hat\d_0$ on $t$ dependent space tensors by
(${\cal L}_\b$ is the Lie derivative with respect to $\b$)
$$ \hat\d_0 \equiv {\d\over \d t} -{\cal L}_\b. $$
The extrinsic curvature $K$ of the space manifold is
$$K_{ij} \equiv -{1\over 2} N^{-1} \hat\d_0 g_{ij}. $$
The \underline{Ricci curvature} of space time admits the 3+1
decomposition, with $H=K^{h}\vb_{h}$
$$R_{ij}= -N^{-1} \hat\d_0 K_{ij} + H K_{ij} - 2K_{im} K^{m}\vb_{j}
- N^{-1} \bar\nabla_j\d_i N + \bar R_{ij},$$
$$R^{0}\vb_j=N^{-1}(\bar\nabla_h K^{h}\vb_{j} - \d_j H), $$
$$R^{0}\vb_{0}= -(N^{-1} \bar\nabla^i \d_i N - K_{ij}K^{ij} + N^{-1}\d_0
H),$$
where an overbar denotes a quantity relative to the space metric $\bar
g\equiv (g_{ij})$.

\section{SECOND ORDER EQUATION FOR $K$}

In the formula above $R_{ij}$, like the right hand side giving its
decomposition, is a $t$ dependent space tensor, the projection on space of
the Ricci tensor of $g$.  We compute its $\hat\d_0$ derivative.  First
we compute $\hat\d_0 \bar R_{ij}$.  The infinitesimal variation of the
Ricci curvature corresponding to an infinitesimal $\delta \bar g$
variation of the space metric is, with $(ij)=ij+ji$ (no factor of $1/2$)
$$ \delta \bar R_{ij}= {1\over 2} \{ \bar\nabla^{h}\bar\nabla_{(i} \delta
g_{j)h} - \bar\nabla_{h}\bar\nabla^{h} \delta g_{ij}
-\bar\nabla_{j}\d_{i}(g^{hk} \delta g_{hk})\}. $$
This expression applies to $(\d/\d t) \bar R_{ij}$ with $\delta
g_{ij} = (\d/\d t) g_{ij}$ and to ${\cal L}_\b \bar R_{ij}$
with $\delta g_{ij} ={\cal L}_\b g_{ij}$.  Therefore, using the
relation between $\hat\d_0 g_{ij}$ and $K_{ij}$, we obtain
\begin{eqnarray*}
\hat\d_0 \bar R_{ij} &=& - \bar\nabla^h \bar\nabla_{(i}(NK_{j)h}) +
\bar\nabla_{h}\bar\nabla^{h} (N K_{ij}) + \bar\nabla_{j}\d_{i}(NH) \\
&\equiv& -\bar\nabla_{(i}\bar\nabla^{h} (NK_{j)h})
+\bar\nabla_{h}\bar\nabla^{h} (N K_{ij}) + \bar\nabla_{j}\d_{i}(NH) \\
&&\hspace{1cm}  - 2N \bar R^{h}\vb_{ijm} K^{m}\vb_h
- N \bar R_{m(i} K_{j)}\vb^m .
\end{eqnarray*}
We now use the expressions for $R^{0}\vb_i$ and $R_{ij}$ to obtain the
identity
\begin{eqnarray*}
\Omega_{ij} &\equiv& \hat\d_0 R_{ij} + \bar\nabla_{(i}(N^2 R_{j)}\vb^0 )
\equiv \\
&&  -\hat\d_0 (N^{-1} \hat\d_0 K_{ij}) +
\bar\nabla^{h} \bar\nabla_{h} (NK_{ij}) + \hat\d_0 (H K_{ij} - 2K_{im}
K^{m}\vb_{j})  \\
&&\hspace{1cm}-\hat\d_0 ( N^{-1} \bar\nabla_j \d_i N) - N \bar\nabla_i\d_j H
\\
&&-\bar\nabla_{(i} ( K_{j)h} \d^h N) - 2N \bar R^{h}\vb_{ijm} K^{m}\vb_{h}
-N\bar R_{m(i} K_{j)}\vb^{m} + H \bar\nabla_j\d_i N .
\end{eqnarray*}

\section{HYPERBOLIC SYSTEM FOR $\protect{\bar g}$, $K$, $N$}

We shall eliminate at the same time the third derivatives of $N$ and the
second derivatives of $H$ as follows (cf. this elimination in the case of
zero shift in \cite{CBR}).  We compute
$$
\hat\d_0 \bar\nabla_j \d_i N \equiv ({\d \over \d t} - {\cal L}_\b)
\bar\nabla_j \d_i N .
$$
We find at once
$$
{\d \over \d t}\bar\nabla_j \d_i N \equiv \bar\nabla_j \d_i {\d N\over \d t}
-{1\over 2} g^{kl}( \bar\nabla_{(i}{\d \over \d t} g_{j)l} -
\bar\nabla_{l}{\d \over \d t} g_{ij}) \d_k N,
$$
and we find an analogous formula when the operator $\d/\d t$ is replaced
by ${\cal L}_\b$.  Finally
$$
\hat\d_0 \bar\nabla_j \d_i N \equiv \bar\nabla_j \d_i \hat\d_0 N +
\{\bar\nabla_{(i} (N K_{j)l}) - \bar\nabla_{l} (N K_{ij})\} \d^l N.
$$
The third order terms in $N$ and second order terms in $H$ in the second
order equation for $K$ can therefore be written in terms of
$$
C_{ij}\equiv N^{-1} \bar\nabla_j \d_i (\hat\d_0 N+N^2 H).
$$
We satisfy the condition $C_{ij}=0$ by requiring $N$ to satisfy the
differential equation (note that $\hat\d_0 N\equiv \d_0 N$)
$$
\d_0 N+N^2 H=0 .\leqno({\rm N}')
$$
Using the expression for $H$, the equation ($N^\prime$) reads
$$
\hat\d_0 \log \{N(\det \bar g)^{-1/2}\}=0.
$$
We find, generalizing the result obtained in \cite{CBR} in the case of
zero shift, that the general solution of this equation is
$$
N=\alpha^{-1} \det (\bar g)^{1/2}, \quad\quad \hat \d_0\alpha=0.
$$
The second equation is a linear differential equation for the scalar
density $\alpha$, depending only on $\b$.  The above choice of $N$ is
called an \underline{algebraic gauge}.  A possible choice if $\b$ does
not depend on $t$ is to take $\alpha$ independent of $t$ and such that
${\cal L}_\b \alpha=0$.  In practice, one may simply regard (N$'$) as a
differential equation to be solved simultaneously with (K$'$) below.

Taking into account the equation (N$'$) satsified by $N$ we see that the
Einstein equations $R_{\alpha \b}=\rho_{\alpha \b}$ imply the wave
equation for $K$
$$
N\Box K_{ij}= Q_{ij} + \Theta_{ij},\leqno({\rm K}')
$$
where we now set
$$
\Box K_{ij} = -N^{-2}\hat\d_0 \hat\d_0 K_{ij} + \bar\nabla^h\bar\nabla_h
K_{ij},
$$
\begin{eqnarray*}
 Q_{ij} &\equiv& - K_{ij} \d_0 H + 2 g^{hm} K_{m(i} \hat\d_0 K_{j)h}
+4 N g^{hl} g^{mk} K_{lk} K_{im} K_{jh} \\
&& + (2\bar\nabla_{(i} K_{j)l}) \d^l N
- 2H N^{-1} \d_i N \d_j N -2\d_{(i}N \d_{j)}H -3\d_h N \bar\nabla^h K_{ij}\\
&&\hspace{0cm}
- K_{ij} \bar\nabla^h \bar\nabla_h N
-N^{-1} K_{ij} (\bar\nabla^h N) \d_h N + N^{-1} K_{h(i} \bar\nabla_{j)}N
\d^h N \\
&& +(\bar\nabla_{(i}\d^h N) K_{j)h}
+ 2N \bar R^{h}\vb_{ijm}K^{m}\vb_{h}
+N\bar R_{m(i} K_{j)}\vb^m - 2 H \bar\nabla_j \d_i N ,
\end{eqnarray*}
$$
\Theta_{ij} \equiv \hat\d_0 \rho_{ij} +\bar\nabla_{(i}(N^2\rho_{j)}\vb^0 ).
$$
An immediate consequence is the following theorem.

{\bf Theorem} {\it In algebraic gauge and with the definition
$K_{ij} \equiv - (2N)^{-1}
\hat\d_0 g_{ij}$, the system $\Omega_{ij}-\Theta_{ij}=0$ is a
quasi-diagonal third order system for $g_{ij}$ with principal operator
$\Box\d_0$.  This system is hyperbolic if $\bar g$ is properly riemannian
and $N^2>0$. (In practice, one can regard the definition of $K_{ij}$ as
an equation ({\rm g}$'$) for $\hat\d_0 g_{ij}$, and then consider the
differential equations ({\rm N}$'$), ({\rm K}$'$) and ({\rm g}$'$) to be
solved simultaneously.)}

The usual local in time existence theorem for a solution of the Cauchy
problem results from the Leray theory of hyperbolic systems\cite{Ler}.

\section{VERIFICATION OF THE ORIGINAL EINSTEIN EQUATIONS}

We set
$$
\Sigma_{\alpha \beta} \equiv (R_{\alpha \beta} -\rho_{\alpha \beta} )
		- {1 \over 2} g_{\alpha \beta} (R- \rho)
{}.
$$
Suppose we have solved the equations
\beq
\Omega_{ij} \equiv \hat \partial_0 R_{ij} + \bar \nabla_{(i}(N^2 R^0_{j)})
=\Theta_{ij},
\eeq
that is, equivalently, doing a few computations,
\beq
\nabla_0 \Sigma_{ij}+\nabla_{(i}(N^2 \Sigma^0_{j)} ) -g_{ij}
\nabla_h(N^2 \Sigma^{0h})+f_{ij}=0,
\eeq
where $f_{ij}$ is linear and homogeneous in $\Sigma_{\alpha \beta}$.

We suppose that the stress energy tensor
$T_{\alpha \beta} \equiv \rho_{\alpha \beta} - {1 \over 2}
g_{\alpha \beta} \rho$
satisfies the conservation laws $\nabla_\alpha T^{\alpha \beta} = 0$. We
deduce then from the Bianchi identities the equations
\beq
\nabla^\alpha \Sigma_{\alpha \beta} = 0.
\eeq

\From\ (4.2) and (4.3) we obtain for $\Sigma_{\alpha \beta}$
a quasi diagonal third order system with principal operator the
hyperbolic operator $\hat \partial_0 \Box$.  The vanishing for $t=0$
of $\Sigma_{\alpha \beta}$ results from the constraints satisfied by
the original initial data
$$
\Sigma^0_j|_{t=0} = 0, ~~~~~~~~~~~~~~~~~\Sigma^{00}|_{t=0} = 0,
$$
and the determination of $\hat \partial_0 K|_{t=0}$ by the equation
$$
\Sigma_{ij}|_{t=0} = 0.
$$
We then deduce from (4.2) and (4.3) the vanishing for $t=0$ of
the derivatives of order $\leq 2$ of $\Sigma_{\alpha \beta}$.
It results therefore, from the uniqueness theorem of Leray[4]
for hyperbolic systems, that a solution of (4.1) with initial data
satisfying the constraints satisfies the full Einstein equations
$\Sigma_{\alpha \beta}=0$.

\section{FIRST ORDER SYSTEM (vacuum)}

The preceding results can be extended without major change to dimensions
greater than 4.  We will show that in dimension 4 a solution of the
vacuum Einstein equations, together with the gauge condition ($N^\prime$),
satisfies a first order symmetric system, hyperbolic if $\bar g$ is
properly Riemannian and $N^2> 0$.  Such a system could be useful to
establish {\it a priori} estimates relevant to global problems, as well
as to the physical applications mentioned in the Introduction.

We have obtained for the unknowns $\bar g$, $K$, and $N$ the equations
\beq
\label{d0g}
\hat\d_0 g_{ij} = -2N K_{ij},
\eeq
\beq
\label{d0N2}
\hat\d_0 (N^2) = -2N H (N^2),
\eeq
\beq
\label{boxK}
N \Box K_{ij} = Q_{ij}.
\eeq
To obtain a first order system we
take as additional unknowns:
$$ N^{-1}\hat\d_0 K_{ij} \equiv L_{ij},\quad \bar\nabla_h K_{ij} \equiv
M_{hij},\quad \d_i \log N \equiv a_i, $$
$$ N^{-1} \hat\d_0 \d_i \log N \equiv a_{0i},\quad a_{ij}=\bar\nabla_i
\d_j log N.$$
We take as equation (\ref{boxK}$'$)
$$
\hat\d_0 K_{ij} = N L_{ij}. \eqno(\ref{boxK}')
$$
The equation (\ref{boxK}) gives
\beq
\label{d0L}
\hat\d_0 L_{ij} -N \bar\nabla^h M_{hij} = N( H L_{ij}-Q_{ij}).
\eeq
In three space dimensions, the Riemann tensor is a linear function of the
Ricci tensor.  Using the equation $R_{ij}=0$ to express $\bar R_{ij}$, we
write $Q_{ij}$ as a polynomial in the unknowns and in $g^{ij}$. ($L_{ij}$
is the essential piece of the spacetime Riemann tensor $R^{\rm o}\vb_{ioj}$.)

We have the following lemma proved for instance by using the
commutativity $\hat\d_0 \d_i= \d_i \hat\d_0$ on components of tensors

\noindent {\bf Lemma} {\it For an arbitrary covariant vector $u_i$ we have}
$$\hat\d_0 \bar\nabla_h u_i = \bar\nabla_h \hat\d_0 u_i + u_l \{
\bar\nabla_{(h}(NK_{i)}\vb^l) -\bar\nabla^l(N K_{ih}) \}.$$
and an analogous formula for tensors with additional terms for each index.

Using this lemma we see that $L_{ij}$ and $M_{hij}$ must satisfy the
equation
\beqa
\label{d0M}
\hat\d_0 M_{kij} -N \overline\nabla_{k} L_{ij} &=&
N\biggl( a_k L_{ij} + M_{k(i}\vb^{l} K_{j)l} + K_{l(i} M_{j)k}\vb^{l}
-K_{l(i} M^{l}\vb_{j)k}   \\
&&\hspace{-1cm} + K_{l(i} ( K^{l}\vb_{j)}a_k + a_{j)} K^{l}\vb_{k}
-a^{l} K_{j)k}) \biggr). \nonumber
\eeqa
On the other hand, (\ref{d0N2}) implies
\beq
\label{d0ai}
\hat\d_0 a_i =-N(H a_i + M_{ik}\vb^{k})
\eeq
while (\ref{d0N2}) and the lemma yield
\beq
\label{d0aji}
\hat\d_0 a_{ji} -N \overline\nabla_{j} a_{i0} = Na_{l} \biggl(
M_{(ij)}\vb^{l}-M^{l}\vb_{ij}
+ a_{(i} K_{j)}\vb^{l} - a^{l} K_{ij}
\biggr) + N a_j a_{i0}.
\eeq
Now we deduce the value of $\hat\d_0 a_{0i}$ from (\ref{d0N2}) and the
Einstein equation
$$ R^0\vb_0 \equiv -\{ N^{-1} \bar\nabla^h \bar\nabla_h N - K_{ij} K^{ij}
+ N^{-1} \d_0 H \} = 0. $$
Indeed these two equations imply
$$\d_0 \d_0 N= N^2 \bar\nabla^h \bar\nabla_h N - N^3 K_{ij} K^{ij} + 2
N^3 H^2.$$
Hence, by differentiation and use of the Ricci formula and the
definitions of $a_{hi}$ and $a_{0i}$, we obtain
\beqa
\label{d0a0i}
\hat\d_0 a_{0i} - N\overline\nabla^{k} a_{ki} &=&
N \biggl(-\overline R^{k}\vb_{i} a_{k} +a_{i}( H^2 -2 K_{kl}K^{kl}
+2 a^k a_k +2 a^{k}\vb_{k} ) \\
&&\vspace{-2cm}+2 a_k a^{k}\vb_{i} + H M_{ik}\vb^{k}
- 2 K^{kl} M_{ikl}\biggr)  \nonumber
{}.
\eeqa
We use again the equation $R_{ij}=0$ in (\ref{d0a0i})
to replace $\bar R_{ij}$ by its
values in terms of the unknowns.

The right hand sides of the equations (\ref{d0g}), (\ref{d0N2}),
(\ref{boxK}$'$), (\ref{d0L}), (\ref{d0M}), (\ref{d0ai}), (\ref{d0aji}),
and (\ref{d0a0i}) are polynomial in the unknowns and $g^{ij}$; they do not
depend on their derivatives.  We eliminate the first derivatives of $\bar
g$ appearing on the left hand sides for instance by introducing on $M$ an
{\it a priori} given metric $e$ which may depend on $t$ but is such that
$$ \hat\d_0 e_{ij}=0.$$
We denote by $D$ the covariant derivative in the metric $e$.  We deduce
then from (\ref{d0g}) the following equation for $G_{hij}\equiv D_h
g_{ij}$:
\beq
\hat\d_0 G_{hij} = -2N \{ a_h K_{ij} + M_{hij} + S^m\vb_{h(i} K_{j)m} \}.
\eeq
The tensor $S$, the difference of the connections of $\bar g$ and $e$, is
given by
$$ S^m\vb_{ij}= {1\over 2} g^{mh} (G_{(ij)h} -G_{hij}) .$$
We thus obtain a \underline{first order differential system}, equivalent
to a \underline{symmetric} system, \underline{hyperbolic if $N^2> 0$ and
$g_{ij}$ is properly riemannian.  Its characteristics}
\underline{are the physical light cone and the ``time'' axis $\d_0$.}

\section{MIXED HYPERBOLIC ELLIPTIC SYSTEM FOR $K$, \protect{$\bar g$},
$N$ WHEN $H$ IS GIVEN}

Another procedure to reduce the second order equation for $K$ obtained above
to a quasi diagonal system with principal part the wave operator is to replace
in the term $\bar \nabla_i \partial_j H$ the mean curvature $H$ by an a priori
given function $h$; this method was used in [3], with $h=0$, in the
asymptotically
euclidean case.  With this replacement the equations $\Omega_{ij}=\Theta_{ij}$
take the form
\beq
N \Box K_{ij}=P_{ij}+\Theta_{ij}
\eeq
where $P_{ij}$ depends on $K$ and its first derivatives and on $\bar g$,
$N$ and $\partial_0 N$ together with their space derivatives of order $\leq 2$.
When $N$ and $\Theta_{ij}$ are known the above equations together with
(5.1) are again a third order quasi-diagonal system for $\bar g$, hyperbolic if
$N>0$ and $\bar g$ properly riemannian.
On the other hand, the equation $R^0\vb_0=\rho^0\vb_0$ together with
$H=h$ implies the equation
\beq
\bar \nabla^i\partial_i N = (K_{ij}K^{ij}-\rho^0\vb_0)N=-\partial_0 h.
\eeq

This equation is an elliptic equation for N when $\bar g$, $K$ and $\rho$
are known.  Note that for energy sources satisfying the ``strong''
energy condition we have $- \rho^0\vb_0 \geq 0$ as well as
$|K|^2\equiv K_{ij}K^{ij} \geq 0$.  These properties are important
for the solution of the elliptic equation.

The hyperbolic system that we have constructed (for given $N$ and $\Theta$)
has local in time solutions for Cauchy data in local Sobolev spaces, like
the one obtained in algebraic gauge.  But now $N$ is determined by an elliptic
equation which has to be solved globally on each space slice $M_t$ at
time $t$.  An iteration procedure leading to a local in time solution has been
used in the vacuum asymptotically flat case with $h=0$ [3].  We give below a
theorem
which applies to the case of a compact $M$ and which can also be proven
by iteration. (Further theorems with $h \neq 0$  on asymptotically
flat spaces can also be given.)

Theorem.  Let $(M,e)$ be a given smooth compact riemannian manifold.
Let there be given on $M\times I$, $I\equiv [0,T]$, a ``pure space''
smooth vector field $\beta$ and a function $h$ such that ($H_k$
is the usual Sobolev space relative to $(M,e)$)
$$
h \in \bigcap_{2\leq k\leq 3} C^{3-k}(I,H_k),~~ \partial_0 h\geq 0,~~
\partial_0 h \not\equiv 0,~~
 h(0,.)\geq 0, {\rm or~~} h(0,.)\leq 0,~~ h \not\equiv 0.
$$

There exists an interval $J\equiv[0,\ell]$, $\ell\leq T$ such
that the system (6.1), with $\Theta_{ij}=0$, (5.1), (6.2)
with $\rho^0\vb_0=0$, has one and only one solution on
$M\times J$,
$$
\bar g \in \bigcap_{1\leq k\leq3} C^{3-k}(J,H_k),~~ K\in
\bigcap_{0\leq k\leq 2}C^{2-k}(J,H_k),~~ N\in \bigcap_{0\leq k\leq2}
C^{2-k}(J,H_{2+k})
$$
with $N>0$ and $\bar g$ uniformly equivalent to $e$,
taking the initial data
$$
g_{ij}(0,.)=\gamma_{ij} \in H_3,~~ K_{ij}(0,.)=k_{ij}\in H_2
$$
if $\gamma$ is a properly riemannian metric uniformly equivalent
to $e$ and $k.k \not\equiv 0$.

It can be proved by using the Bianchi identities that if the
initial data satisfy the constraints the constructed solution
satisfies the original Einstein equations.

\vspace{1cm}
J.W.Y. acknowledges support from the National Science Foundation of the
U.S.A., grants PHY-9413207 and ASC/PHY 93-18152 (DARPA supplemented).

\end{document}